\documentclass[10pt, conference, compsocconf]{IEEEtran}
\usepackage{footnote}
\usepackage{subfigure}
\usepackage{float}
\usepackage{multirow}
\usepackage{graphicx}
\usepackage[cmex10]{amsmath}
\usepackage{amsmath}
\usepackage{amssymb}
\usepackage{algorithm}
\usepackage{algorithmic}
\usepackage{xcolor}

\def\IEEEbibitemsep{0pt plus .5pt}

\begin{document}

\bstctlcite{IEEEexample:BSTcontrol}
%
% paper title
% can use linebreaks \\ within to get better formatting as desired
\title{Robust Composition of Drone Delivery Services under Uncertainty}
% \vspace{-4cm}

% author names and affiliations
% use a multiple column layout for up to two different
% affiliations

%\author{\IEEEauthorblockN{Babar Shahzaad, Athman Bouguettaya, Sajib Mistry, Azadeh Ghari Neiat }
%\IEEEauthorblockA{School of  Computer Science,\\
%The University of Sydney, Australia\\
%\{babar.shahzaad, athman.bouguettaya, sajib.mistry, azadeh.gharineiat \}@sydney.edu.au\\
%}
%}

\author{\IEEEauthorblockN{Babar Shahzaad\IEEEauthorrefmark{1},
Athman Bouguettaya\IEEEauthorrefmark{1},
Sajib Mistry\IEEEauthorrefmark{2}
}

\IEEEauthorblockA{\IEEEauthorrefmark{1}School of  Computer Science,
The University of Sydney, Australia\\
\{babar.shahzaad, athman.bouguettaya\}@sydney.edu.au}
\IEEEauthorblockA{\IEEEauthorrefmark{2}School of Electrical Engineering, Computing and Mathematical Sciences,
Curtin University, Australia\\
sajib.mistry@curtin.edu.au
}
}

% conference papers do not typically use \thanks and this command
% is locked out in conference mode. If really needed, such as for
% the acknowledgment of grants, issue a \IEEEoverridecommandlockouts
% after \documentclass

% for over three affiliations, or if they all won't fit within the width
% of the page, use this alternative format:
% 
%\author{\IEEEauthorblockN{Michael Shell\IEEEauthorrefmark{1},
%Homer Simpson\IEEEauthorrefmark{2},
%James Kirk\IEEEauthorrefmark{3}, 
%Montgomery Scott\IEEEauthorrefmark{3} and
%Eldon Tyrell\IEEEauthorrefmark{4}}
%\IEEEauthorblockA{\IEEEauthorrefmark{1}School of Electrical and Computer Engineering\\
%Georgia Institute of Technology,
%Atlanta, Georgia 30332--0250\\ Email: see http://www.michaelshell.org/contact.html}
%\IEEEauthorblockA{\IEEEauthorrefmark{2}Twentieth Century Fox, Springfield, USA\\
%Email: homer@thesimpsons.com}
%\IEEEauthorblockA{\IEEEauthorrefmark{3}Starfleet Academy, San Francisco, California 96678-2391\\
%Telephone: (800) 555--1212, Fax: (888) 555--1212}
%\IEEEauthorblockA{\IEEEauthorrefmark{4}Tyrell Inc., 123 Replicant Street, Los Angeles, California 90210--4321}}

% use for special paper notices
%\IEEEspecialpapernotice{(Invited Paper)}

% make the title area
\maketitle

\begin{abstract}
We propose a novel robust composition framework for drone delivery services considering changes in the wind patterns in urban areas. The proposed framework incorporates the dynamic arrival of drone services at the recharging stations. We propose a Probabilistic Forward Search (PFS) algorithm to select and compose the best drone delivery services under uncertainty. A set of experiments with a real drone dataset is conducted to illustrate the effectiveness and efficiency of the proposed approach.

\end{abstract}
% \vspace{-2mm}

% For peer review papers, you can put extra information on the cover
% page as needed:
% \IFCLASSOPTIONpeerreview
% \begin{center} \bfseries EDICS Category: 3-BBND \end{center}
% \fi
%
% For peerreview papers, this IEEEtran command inserts a page break and
% creates the second title. It will be ignored for other modes.
\IEEEpeerreviewmaketitle

% \vspace{-4mm}
\section{Introduction}
% no \IEEEPARstart

Drones have created a myriad of new opportunities in various practical fields \cite{DBLP:journals/corr/abs-1805-00881}.
Commercial applications of drones include agriculture, healthcare, and delivery \cite{9284115}. During the COVID-19 pandemic, several countries have leveraged drone technology to provide safe, contactless, and more resilient alternatives to deliver goods in remote locations \cite{9086010}. Major competitors in the delivery service industry such as Amazon and Google are expanding the use of drones for delivery services as a complement to traditional delivery modes \cite{doi:10.1111/drev.10313}. Drones offer \emph{fast}, \emph{convenient}, and \emph{cost-effective} delivery services compared to land-based delivery \cite{5}.

The \textit{service paradigm} \cite{Bouguettaya:2017:SCM:3069398.2983528} provides a powerful abstraction of the \emph{functional} and \emph{non-functional} or \emph{Quality of Service} (QoS) properties of a drone termed as \textit{Drone-as-a-Service} (DaaS) \cite{8818436}. The functional property is expressed as the package delivery from a designated source (take-off) station (e.g., warehouse rooftop) to a destination (landing) station. The QoS properties of a DaaS include payload, battery capacity, and flight range. DaaS usually uses a \textit{skyway} network to operate in a geographic area \cite{west2015drone}. A skyway network is defined as joining a set of nodes (vertices) representing take-off and/or landing and/or recharging stations. A line segment between any two nodes represents the service abstraction. An instantiation of this service representation is delivering a package between the two nodes of the segment under a set of requirements/constraints.

\textit{DaaS composition} is the process of selecting the best drone-based services that form a skyway path from a given source to a destination \cite{10.1007/978-3-030-33702-5_28}. The composition is naturally fit to deliver packages considering different QoS requirements from the end-users, e.g., fastest, cost-efficient, safe, and contactless delivery.
 
Existing DaaS composition approaches are deterministic in nature \cite{10.1007/978-3-030-33702-5_28, 9284135, alkouz2020formationbased}. These compositions do not consider uncertainties that affect the QoS of the composition, e.g., failed deliveries, longer delivery time, and excessive battery recharging time at intermediate nodes. \emph{Uncertainty} is an extrinsic part of the dynamic drone service environment. For example, uncertainty in the flight behavior is caused by real-time variations in environmental conditions such as wind conditions and temperature \cite{10}. High-rise buildings greatly impact the amplification of wind speed and temperature in urban areas \cite{en11092204}. In this study, \emph{our particular focus is on the wind effects on drone delivery time, including ``headwind" and ``tailwind"}. Flying with strong wind could increase (tailwind) or decrease (headwind) energy consumption and drone speed \cite{Citroni_2019}. The headwind reduces the flight range of a drone while tailwind extends the flight range. The accurate predictions of wind for a long-term period may not be possible due to its highly stochastic nature \cite{7990193}. In addition, multiple drone delivery services which operate in the same network may cause \emph{congestion} at certain charging stations, thus creating uncertainty. Congestion occurs due to the simultaneous arrival of other drones at the same station occupying all the available pads.

\emph{We propose a robust DaaS composition framework to select and compose a set of best drone services}. In this paper, the term robust refers to the algorithm's ability to resist the unwanted effects of uncertain environmental changes. We assume that no handover of packages occurs among drones in the air or at intermediate stations as each drone has its own delivery plan, i.e., the same drone delivers the package from source to destination. We focus on recharging constraints and changes in wind patterns as uncertainty factors in urban settings. We assume that intrinsic factors are deterministic, whereas extrinsic factors are stochastic. The payload, battery capacity, and flight range of each drone are known beforehand. The availability of recharging pads at each station and the wind conditions are assumed to be \emph{probabilistic in nature}. 

We propose a Probabilistic Forward Search (PFS) algorithm to find the robust DaaS composition. The PFS approach uses a probabilistic distribution of predicted wind for the selection of drone services. The decision-making in the PFS approach is based on the recharging pads' availability at adjacent and next-to-adjacent recharging stations. In addition, the decision-making depends upon the wind conditions on the skyway segments leading to these recharging stations. Therefore, the PFS approach may provide better results than baseline approaches. The main contributions of this paper are as follows:
\begin{itemize}
    \item[$\bullet$] Designing an uncertainty-aware DaaS system model for drone-based services.
    \item[$\bullet$] Developing a Probabilistic Forward Search (PFS)-based approach for the best drone service selection and composition.
    \item[$\bullet$] Conducting experiments using a real drone dataset to illustrate the performance of the proposed probabilistic composition approach.
\end{itemize}

\section{Related Work} \label{relatedwork}

In recent years, the routing and scheduling of drone services has become an active area of research. Most existing works focus on combined delivery of drones with ground vehicles. An adaptive large neighbourhood search method is developed to address the TSP with multiple drones \cite{10.1145/3287921.3287932}. The goal of this work is to reduce the delivery cost to serve all customers for both ground vehicle and drones. A greedy algorithm is used for generating an initial TSP solution. The proposed approach uses the initial solution to remove nodes from the vehicle route and adds them to the drone routes. A TSP solution with multiple drones is compared to a TSP solution with a single drone. It is concluded that more drones support the generation of efficient routes in comparison to a single drone. The proposed method uses a ground vehicle for deliveries \textit{which is not appropriate for locations with no road access}. In addition, the proposed method \textit{does not take into account the recharging constraints} and \textit{wind uncertainty}.

An energy consumption model is presented for automated drone delivery services in \cite{choi2017optimization}. They assumed that drones can perform multi-package deliveries in a predefined service area. The drone fleet size is optimized by analyzing the impact of payload weight and flight range considering battery capacity. They explore the relationship between four variables (working period, drone speed, demand density of service area, and battery capacity) to minimize the total costs of the drone delivery system. The study indicated that the long hours of operation would benefit both service providers and customers. They found that drone deliveries are more cost-effective in areas with high demand densities. \textit{This study does not consider the dynamic congestion conditions at recharging stations and uncertain wind conditions}.

There is a paucity of literature concerning the wind effects on drone delivery. Selecky et al. \cite{selecky2013wind} studied the wind effects which influences the flight direction of a drone. An accelerated A* algorithm is developed to incorporate the wind effects and generate reachable states. The wind is assumed to be constant which does not capture the real-world scenarios. The proposed approach does not consider the wind variation and wind uncertainty in different areas.

The drone delivery problem is abstracted using service paradigm in \cite{8818436}. The function of the drone is to deliver a package from one node to another node over a line segment in the skyway network. A service model is designed considering the spatio-temporal feature of drone services. A heuristic-based algorithm is developed to select and compose right drone services taking into account the QoS properties. The proposed approach focuses only on the deterministic properties of services which is not realistic. This work has been extended in \cite{10.1007/978-3-030-33702-5_28} as a constraint-aware deterministic composition approach to incorporate the recharging constraints at stations. A lookahead heuristic-based algorithm is presented for selection and composition of optimal services.

A resilient composition framework is proposed for drone delivery services considering congestion conditions at recharging stations \cite{SHAHZAAD2021335}. The framework includes a formal service model for the representation of constraint-aware drone services. An initial offline drone service composition plan is generated using a deterministic lookahead algorithm. A heuristic-based resilient composition approach is proposed to adapt the runtime changes in the initial composition plan and update it to meet the delivery requirements of the user. The proposed approach does not consider the probabilistic nature of the wind, which changes with time. In addition, the robustness of the composition is not considered, which is of paramount importance for efficient service delivery. To the best of our knowledge, this paper is the first attempt to present a robust DaaS composition that considers the uncertain wind conditions.

\section{Uncertainty-aware DaaS System Model}
We propose an uncertainty-aware DaaS system model for drone delivery services. The proposed system model is divided into two sub-models: (1) DaaS Model and (2) DaaS Delivery Model under Uncertainty.

\subsection{DaaS Model}
The DaaS, composite DaaS service, and DaaS composition problem are defined as follows.\\
\textbf{Definition 1: Drone-as-a-Service (DaaS)}. A DaaS is a 3-tuple $<DaaS\_id,$ $DaaS_{f}, DaaS_q>$, where
\begin{itemize}
    \item[$\bullet$] $DaaS.id$ is a unique drone service ID,
    \item[$\bullet$] $DaaS_{f}$ represents the delivery function of a drone over a line segment. The location and time of a DaaS are 2-tuples $<loc_s,loc_e>$ and $<t_s, t_e>$, where
    \begin{itemize}
        \item $loc_s$ and $loc_e$ represent the pickup location and the delivery location,
        \item $t_s$ and $t_e$ represent the start time and the end time,
    \end{itemize}
    \item[$\bullet$] $DaaS_q$ is an n-tuple $<q_1, q_2,\ldots,q_n>$, where each $q_i$ represents a quality parameter of a DaaS, e.g., flight range.
\end{itemize}

\textbf{Definition 2: Composite DaaS Service (CS)}. A CS is a 3-tuple $<CSID, CSF, CSQ>$, where
\begin{itemize}
    \item[$\bullet$] $CSID$ is a unique $CS$ identifier,
    \item[$\bullet$] $CSF$ is a set of functions $\{f_1(DaaS_1),$ $f_2(DaaS_2),$ $\ldots,$ $f_n(DaaS_n)\}$, where each $f_i$ represents function of corresponding component DaaS $DaaS_i \in  CS$
    \item[$\bullet$] $CSQ$ is an m-tuple $<Q_1, Q_2, \ldots, Q_m>$, where each $Q_j$ denotes an aggregated value of $j^{th}$ quality parameter of component DaaS $DaaS_i \in CS$.
\end{itemize} 

\textbf{Definition 3: DaaS Composition Problem}. We build upon and extend the drone service and quality model proposed in \cite{8818436}. Given a set of DaaS services $S_{DaaS} = \{DaaS_1, DaaS_2,..., DaaS_n\}$, the DaaS composition problem is to compose the services for delivering a package from the warehouse to the customer location in minimum time.

\subsection{DaaS Delivery Model under Uncertainty}

In the existing DaaS model \cite{8818436}, all the drone services and the service environment are deterministic, i.e., the availability and QoS-values of services are known beforehand. We extend this model by introducing the dynamic recharging constraints and incorporating the wind effects.

The stochastic nature of wind has a significant nonlinear effect on the \emph{energy consumption rate} and \emph{flight range} of a drone \cite{Citroni_2019}. The headwinds drain the drone's energy more quickly, while the tailwinds reduce energy consumption. We determine the impact of wind speed and direction (i.e., headwind and tailwind) on the travel time of the drone. We use the method in \cite{10.2307/43943662} to calculate the effects of headwind and tailwind on the travel time from node $i$ to $j$ as follows.

\begin{align}
\begin{split}\label{eq:1}
    \delta &= \theta_{ij} - \theta_{WS}
\end{split}\\
\begin{split}\label{eq:2}
    A &= WS.\cos(180-\delta)
\end{split}\\
\begin{split}\label{eq:3}
    C &= WS.\sin(180-\delta)
\end{split}\\
\begin{split}\label{eq:4}
    B &= \sqrt{AS^2-C^2}
\end{split}\\
\begin{split}\label{eq:5}
    GS &= A + B \\
    &= WS.\cos(180-\delta) + \sqrt{AS^2-WS^2.\sin^2(180-\delta)}
\end{split}\\
\begin{split}\label{eq:6}
    T_{ij} &= \frac{d_{ij}}{GS} 
\end{split}
\end{align}
where $\theta_{ij}$ is the bearing from node $i$ to $j$, $\theta_{WS}$ is the wind bearing, $\delta$ is the course correction angle, $WS$ is the wind speed, $A$ is the headwind/tailwind, $C$ and $B$ are the wind adjustment angles, $AS$ is the air speed, $GS$ is the ground speed, $d_{ij}$ is the distance between nodes $i$ and $j$, and $T_{ij}$ is the travel time from node $i$ to $j$. When $|\delta| < 90$, $A$ is negative and denotes headwind. When $90 < |\delta| \leq 180$, $A$ is positive and denotes tailwind.

A drone can take a finite set of actions at each node during its journey from source to destination. A drone can either \emph{wait}, \emph{recharge}, or \emph{travel} from one node to another. \emph{The objective is to take the right actions to reach the destination faster}. Based on the probability distribution of wind speed and direction, the actions create a huge set of state space. Therefore, \emph{we transform the problem of action selection at any node into probabilistic state transition tree}. We formally define a state as follows.

\textbf{Definition 4: State}. A state is a 3-tuple $<N.id,$ $TSTP,$ $BP>$, where
\begin{itemize}
    \item[$\bullet$] $N.id$ is a unique node identifier,
    \item[$\bullet$] $TSTP$ is a timestamp which represents a recorded time snapshot of a state,
    \item[$\bullet$] $BP$ represents the number of busy recharging pads at a certain station.
\end{itemize}

A single DaaS may not satisfy a user's long-distance delivery requirements. In such cases, a robust drone service composition is required from a large set of candidate services. The DaaS composition under uncertainty is a challenging task. An adjacent attractive drone service may lead to a highly time-expensive service. For instance, we are given a skyway network with the source location and destination location. Our target is to find the selection and composition of temporally optimal drone services considering wind conditions. In this context, temporally optimal refers to leading towards the destination \emph{faster}. The wind conditions are \emph{time-variant} which cause uncertainty for future drone services. We require to predict the future wind conditions and their effects on the overall delivery time. We focus on the uncertain wind conditions that may result in longer delays for delivery by drones.

\subsection{Informed Exhaustive Search}

Informed Exhaustive Search (IES) approach is an all-paths search method \cite{DBLP:journals/corr/abs-1902-03506}. In this approach, no uncertainty is involved during the service composition process, i.e., the actual information about wind speed and direction is known. IES computes all possible DaaS compositions from source to destination and selects an optimal composition based on QoS parameters (i.e., delivery time). Finding all possible DaaS compositions is computationally not feasible and limits the use of IES for large-scale problems. The time complexity of IES is exponential, which reduces its performance significantly.

\subsection{Robust Composition using Probabilistic Forward Search}
\begin{figure} [t]
    \centering
    \includegraphics[width=0.48\textwidth]{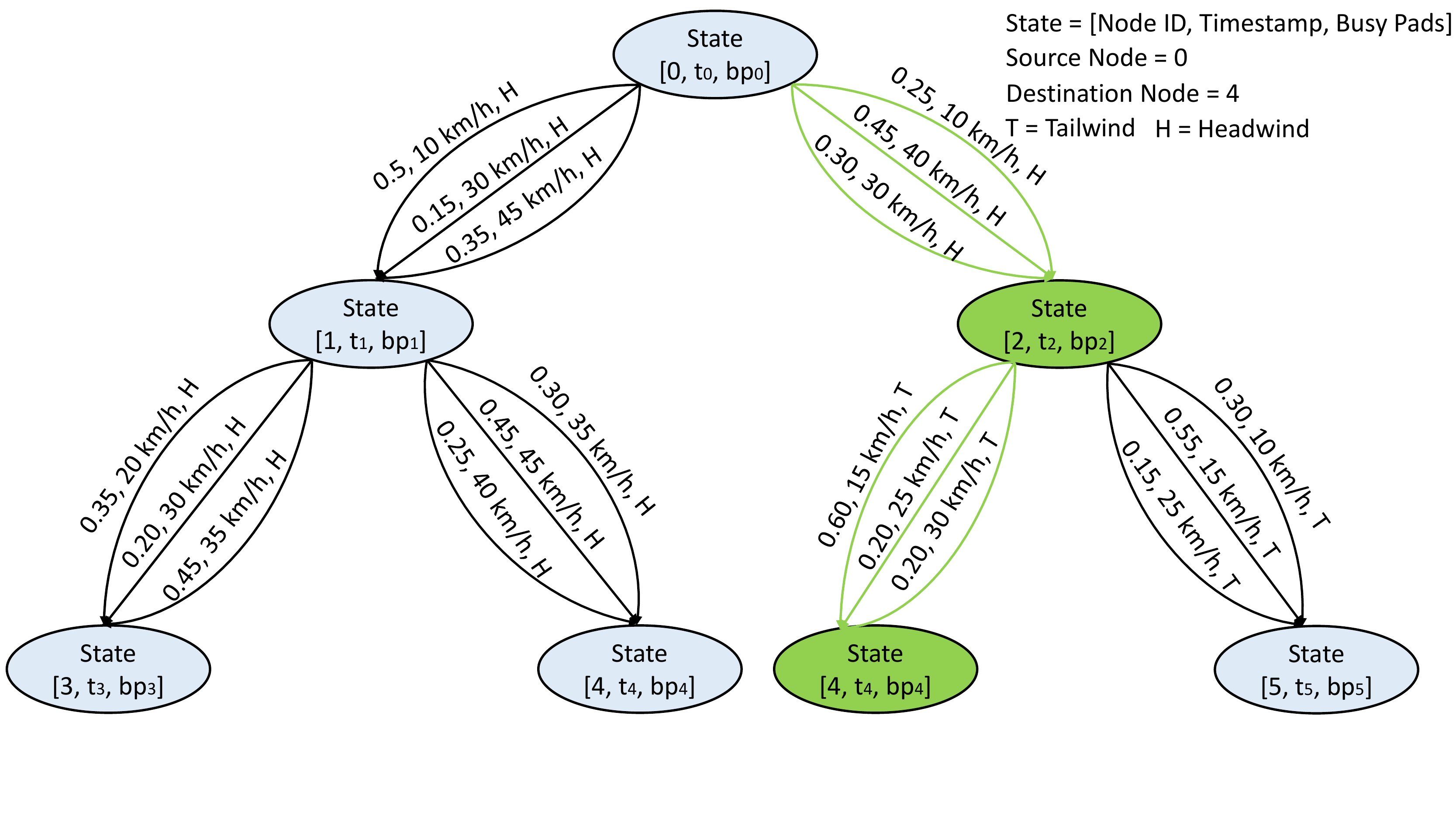}
    \caption{State selection using PFS}
    \label{fig4}
    \vspace{-0.4cm}
\end{figure}
We propose a Probabilistic Forward Search (PFS) heuristic-based solution for robust DaaS composition under uncertainty. In the proposed approach, the term forward search refers to considering next-to-adjacent states (services) in decision making. The accuracy of the predicted probability distribution of wind affects the selection of an optimal drone service. Our predictive model is based on the historical data of wind speed and direction in a specific time slot. The prediction is made for each skyway segment to estimate the arrivals of other drone services in the future. Most of the existing prediction models focus on the shortest path from source to destination. Due to the wind uncertainty and congestion conditions at recharging stations, the shortest path's travel time may not guarantee the overall shortest time from source to destination. We compute the sum of the travel, waiting, and recharging times to select a service. Fig. \ref{fig4} illustrates how the PFS approach favors the optimal service (state) selection. Using the PFS approach provides more information to guide the selection of overall optimal states. The principle of PFS-based tree exploration is similar to a depth-first search. However, we use the PFS approach to select only one service at a time. The details are given in Algorithm \ref{alg:algorithm1}.
\begin{algorithm}[t]
\small
\caption{Probabilistic Forward Search based DaaS Composition}\label{alg:algorithm1}
	\begin{algorithmic}[1]
    \REQUIRE
		Spatio-Temporal Graph $G$, Drones $D$, Source $src$, Destination $dst$, Package Weight $w$, Forward Search $fs$, Start Time $sTime$, Wind $Wi$, Probability Distribution $Pr$
	\ENSURE
		DaaS Composition $CompDaaS$\\
		\STATE $curLoc \gets src$,\quad $curTime \gets sTime$
		\STATE $d_{sel} \gets$ block\_nested\_loop ($D$, $w$)
		\STATE $St_{reach} \gets$ get\_reachable\_stations ($G$, $src$, $d_{sel}$, $w$, $Wi$, $Pr$)
		\IF{$dst \in St_{reach}$}
		\STATE $deliveryTime \gets$ travel\_time ($src$, $dst$, $Wi$)
		\STATE $CompDaaS \gets$ skyway\_segment ($src$, $dst$)
		\RETURN $CompDaaS$
		\ELSE
		\WHILE{$St_{reach} \ne \phi$}
		\IF{$dst \notin St_{reach}$}
		\STATE $nextStTime \gets $ reachable\_station\_time ($G$, $curLoc$, $d_{sel}$, $w$, $Wi$, $St_{reach}$, $fs$, $dst$, $Pr$)
		\STATE $nextSt \gets $ $St_{reach}$ [index (min ($nextStTime$))]
		\STATE $deliveryTime \gets$ ($curTime$ + travel\_time ($curLoc$, $nextSt$, $Wi$)   + wait\_and\_recharge\_time ($G$, $nextSt$))
		\STATE $DaaS \gets$ skyway\_segment ($curLoc$, $nextSt$) 
		\STATE $CompDaaS$.append($DaaS$)
		\STATE $curLoc \gets nextSt$
		\STATE $curTime \gets deliveryTime$
		\STATE $St_{reach} \gets$ get\_reachable\_stations ($G$, $curLoc$, $d_{sel}$, $w$, $Wi$, $Pr$)
		\ELSE
		\STATE $deliveryTime \gets$ ($curTime$ + travel\_time ($curLoc$, $dst$, $Wi$)
		\STATE $DaaS \gets$ skyway\_segment ($curLoc$, $dst$)
		\STATE $CompDaaS$.append ($DaaS$)
		\RETURN $CompDaaS$
		\ENDIF
		\ENDWHILE
		\ENDIF
		\STATE print (``No suitable composition found for the given source to destination")
	\end{algorithmic}
\end{algorithm}

In Algorithm \ref{alg:algorithm1}, the output is a robust DaaS composition from a designated source to a destination. We first use the \emph{Block Nested Loop (BNL)} \cite{6916872} algorithm for selecting a set of optimal drones from a large set of delivery drones given the payload weight (Line 2). Multiple service providers offer drone services. Each provider has several drones with different quality attributes. The BNL approach helps select a set of drones determined to be a good fit for the delivery request. We then obtain a set of reachable stations by the selected drone considering the payload weight and predicted probability distribution of wind (Line 3).
The get\_reachable\_stations function finds the nearby stations based on the travel distance from the current drone position. If the desired destination lies within reachable stations, the drone delivers the package without recharging (Line 4-7). In such a case, service composition is not required (a single drone service fulfills the request). We compute the time to travel the skyway segment from source to destination considering wind conditions (Line 5). We compose optimal drone services from source to destination to serve long-distance areas (Lines 9-25). If the destination does not lie within reachable stations, we compute the time to each reachable station (Line 11).

We select the optimal drone service based on the travel time (calculated using equation \ref{eq:6}), its probability of occurrence, availability of the recharging pads, current waiting time, expected waiting time on the next node, and flight time to the destination (Lines 12-13). On each iteration, we add the selected optimal services to $DaaS\_Comp$  (Line 15). We update the current location and time of the drone (Lines 16-17). This process continues till the destination node is discovered or the reachable stations' list is empty (i.e., no suitable composition found).

\section{Performance Evaluation}
In this section, we analyze the effectiveness of the Probabilistic Forward Search (PFS) approach.

\subsection{Experiment Settings with Real-world Datasets}

We develop a robust DaaS composition framework for delivery services to evaluate the performance of the PFS approach. We build a skyway network using the NetworkX python library, where each node can be a delivery target or a recharging station. We model multiple drone services from different drone service providers operating in the same network. The drone set consists of quality parameters of each drone operating in the skyway network, e.g., flight range and payload capacity. The experiments are conducted for an average of 50\% times the total number of nodes. For example, if there are 40 nodes in the network, the experiment is performed 20 times. We select a random source and a random destination point for each experiment. We use a real dataset of drone trajectories, including data for altitude, coordinates, and timestamps \cite{14}. We augment a dataset for different types of drones considering the payload, speed, flight range, recharging time, and battery capacity. The experimental variables are described in Table \ref{tab:table1}. All the experiments are run on an Intel Core i9-9900X processor (3.50 GHz and 32.0 GB RAM) under Windows 10. All the algorithms are written in Python.

\begin{table}[t]
\centering
\caption{Experimental Variables}
\label{tab:table1}
\begin{tabular}{|l|l|}
\hline
 \textbf{Variable} &  \textbf{Values} \\

\hline

Drone model &  DJI M200 V2\\ \hline

Maximum payload capacity & 1.45 Kg \\ \hline

Maximum drone flight time & 24 min \\ \hline

Maximum drone flight range & 32.4 km \\ \hline

Maximum drone speed & 81 km/h \\ \hline

Recharging time from 0\% to 100\% & 2.24 hours \\ \hline

Maximum nodes in the skyway network &  1000  \\ \hline

No. of pads at each recharging station & 5 \\ \hline

Experiment run the total number of nodes & 50\% \\

\hline
\end{tabular}
\end{table}

\subsection{Results and Discussion}

The PFS approach performs robust composition of the right drone services to deliver the package faster.

\begin{figure}[t]
    \centering
        \centering
        \includegraphics[width=0.4\textwidth]{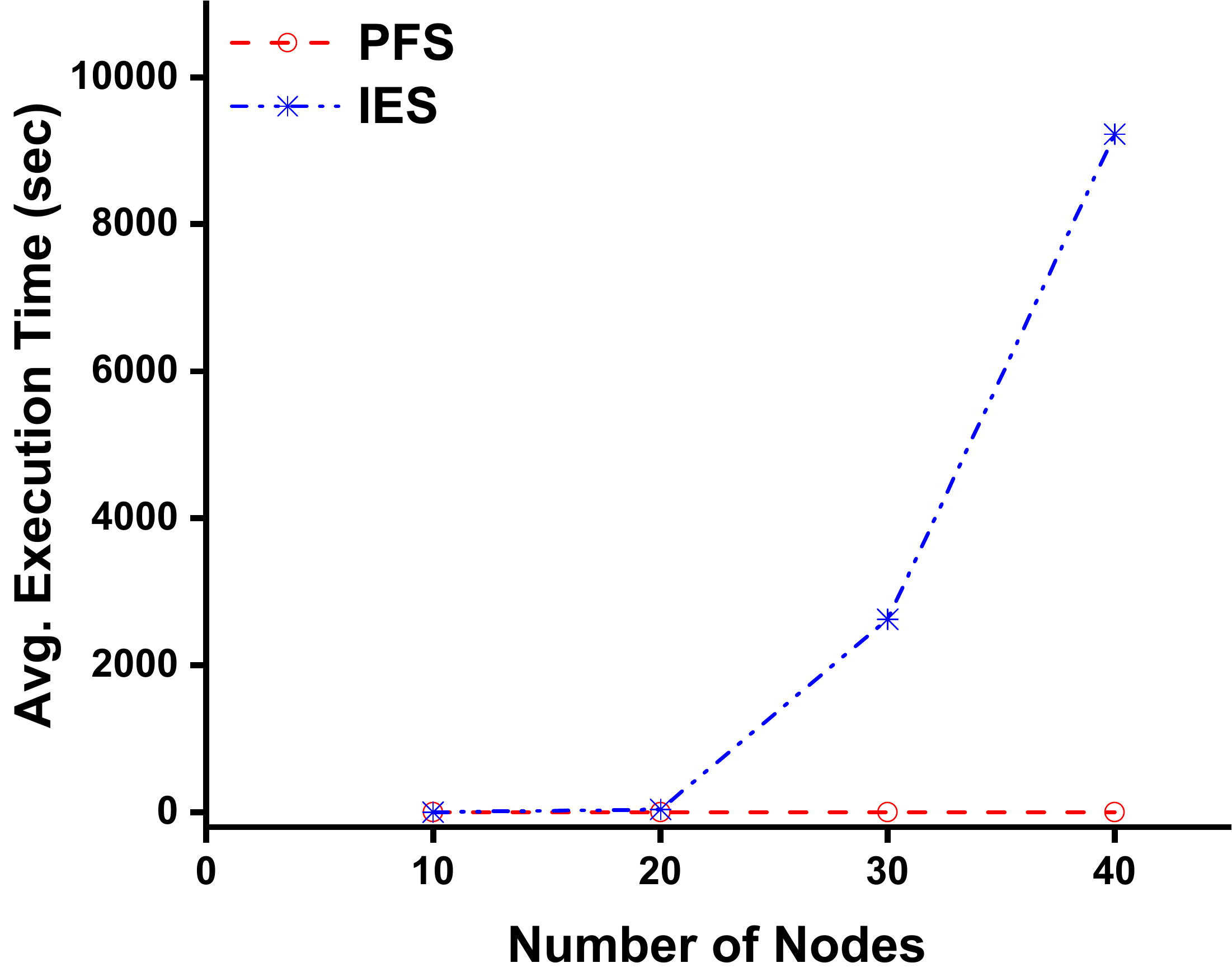} % first figure itself
        \caption{Average execution time}
        \label{fig:fig7.pdf}
    % \vspace{-0.3cm}
\end{figure}

\subsubsection{Average Execution Time}
The time complexity is an important parameter to evaluate the performance of an algorithm. The IES approach is very computationally expensive compared to the proposed PFS approach. The execution time increases as the number of possible drone service compositions increase. The average execution times for IES and PFS approaches are presented in  Fig.~\ref{fig:fig7.pdf}. The execution time for the PFS approach is much less because it avoids exhaustive drone service compositions. As expected, the average execution time for the IES approach grows exponentially for an increasing number of nodes. The experiments indicate that when the nodes are above 40, the results' trends are similar. As a result, we set the maximum number of nodes at 40 for the skyway network. The use of the baseline approach is not practical in real-world scenarios for large-scale problems because of its exhaustive nature.

\subsubsection{Average Delivery Time}
\begin{figure}[t]
    \centering
        \centering
        \includegraphics[width=0.4\textwidth]{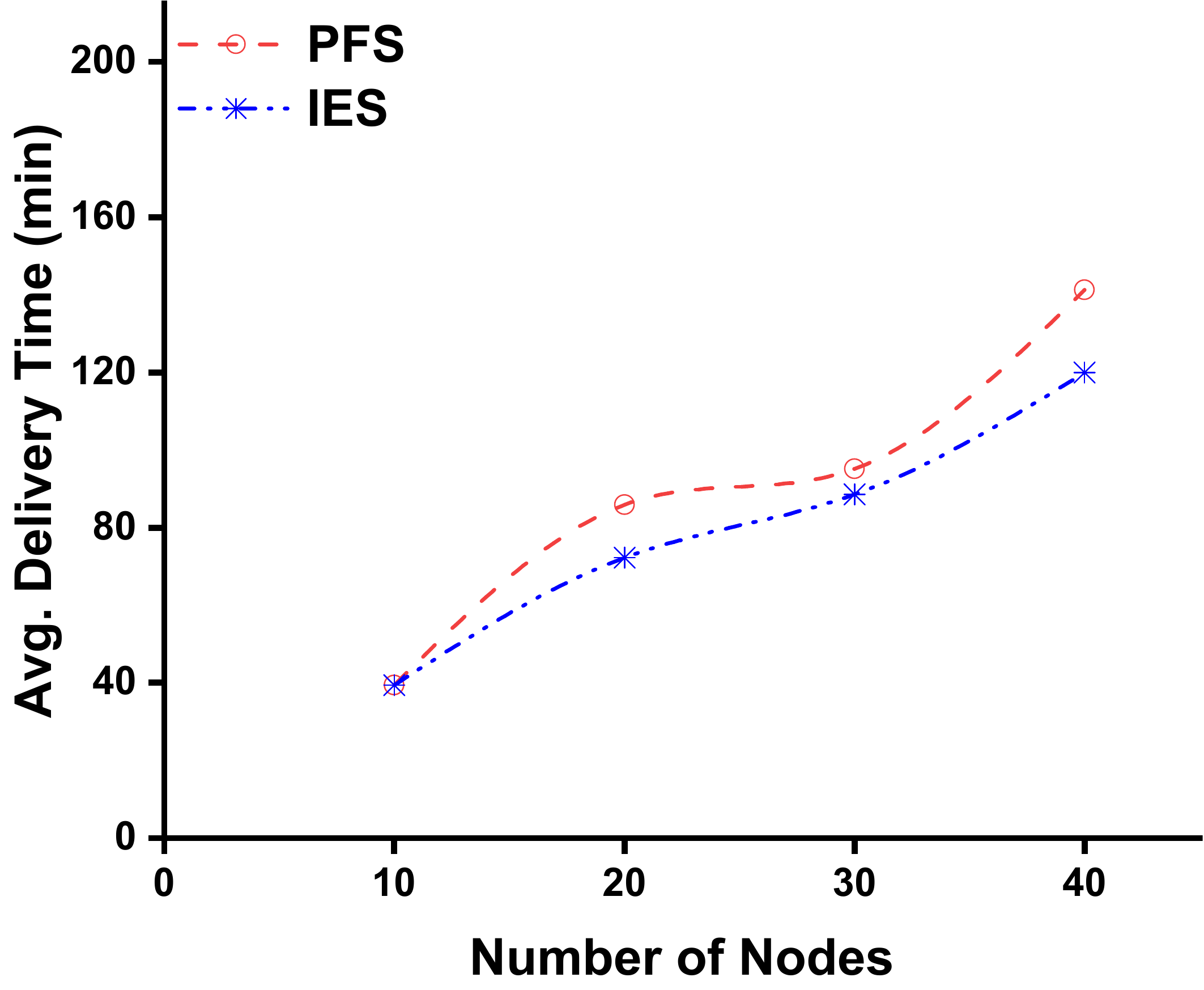} % first figure itself
        \caption{Average delivery time}
        \label{fig:fig5.pdf}
    % \vspace{-0.3cm}
\end{figure}

The delivery time of a drone is a summation of recharging, waiting, and flight times. The delivery time is mainly affected by the occupancy of certain recharging stations for long periods of time. Fig.~\ref{fig:fig5.pdf} shows the delivery times of IES and PFS approaches. The IES approach always computes all possible drone service compositions, which in turn provides exact solutions. The decision-making of the proposed PFS approach relies on the congestion information from the next-to-adjacent nodes. Therefore, the PFS approach provides delivery solutions close to the IES approach. However, the PFS approach is significantly faster than the IES approach, as shown in  Fig.~\ref{fig:fig7.pdf}.

\section{Conclusion}

We propose a novel framework for robust drone service composition considering uncertain wind conditions over the skyway segments in a skyway network. A Block Nested Loop algorithm is used for the selection of the right drone at the source location. The proposed approach incorporates the dynamic arrival of drone services at the recharging stations. We select and compose the optimal services using the Probabilistic Forward Search (PFS) approach to minimize the delivery time. We run a set of experiments to evaluate the efficiency of the proposed approach compared to IES approach. The experimental results prove that the proposed approach is computationally efficient than the IES approach. Moreover, our proposed approach is a practical solution for real-world scenarios of drone delivery services due to its computational efficiency and near-optimal solutions. We plan to consider handover among drones at recharging stations in the future.

\section*{Acknowledgment}
This research was partly made possible by DP160103595 and LE180100158 grants from the Australian Research Council. The statements made herein are solely the responsibility of the authors.

\bibliographystyle{IEEEtran}
\bibliography{references}

% Generated by IEEEtran.bst, version: 1.14 (2015/08/26)
\begin{thebibliography}{10}
\providecommand{\url}[1]{#1}
\csname url@samestyle\endcsname
\providecommand{\newblock}{\relax}
\providecommand{\bibinfo}[2]{#2}
\providecommand{\BIBentrySTDinterwordspacing}{\spaceskip=0pt\relax}
\providecommand{\BIBentryALTinterwordstretchfactor}{4}
\providecommand{\BIBentryALTinterwordspacing}{\spaceskip=\fontdimen2\font plus
\BIBentryALTinterwordstretchfactor\fontdimen3\font minus
  \fontdimen4\font\relax}
\providecommand{\BIBforeignlanguage}[2]{{%
\expandafter\ifx\csname l@#1\endcsname\relax
\typeout{** WARNING: IEEEtran.bst: No hyphenation pattern has been}%
\typeout{** loaded for the language `#1'. Using the pattern for}%
\typeout{** the default language instead.}%
\else
\language=\csname l@#1\endcsname
\fi
#2}}
\providecommand{\BIBdecl}{\relax}
\BIBdecl
\renewcommand{\BIBentryALTinterwordstretchfactor}{4}

\bibitem{DBLP:journals/corr/abs-1805-00881}
H.~Shakhatreh \emph{et~al.}, ``Unmanned aerial vehicles (uavs): A survey on
  civil applications and key research challenges,'' \emph{IEEE Access}, vol.~7,
  pp. 48\,572--48\,634, Apr 2019.

\bibitem{9284115}
B.~Shahzaad \emph{et~al.}, ``A game-theoretic drone-as-a-service composition
  for delivery,'' in \emph{IEEE International Conference on Web Services
  (ICWS)}, 2020, pp. 449--453.

\bibitem{9086010}
V.~{Chamola} \emph{et~al.}, ``A comprehensive review of the covid-19 pandemic
  and the role of iot, drones, ai, blockchain, and 5g in managing its impact,''
  \emph{IEEE Access}, vol.~8, pp. 90\,225--90\,265, 2020.

\bibitem{doi:10.1111/drev.10313}
D.~Bamburry, ``Drones: Designed for product delivery,'' \emph{Design Management
  Review}, vol.~26, no.~1, pp. 40--48, Jul. 2015.

\bibitem{5}
R.~D'Andrea, ``Guest editorial can drones deliver?'' \emph{IEEE Transactions on
  Automation Science and Engineering}, vol.~11, no.~3, pp. 647--648, Jul. 2014.

\bibitem{Bouguettaya:2017:SCM:3069398.2983528}
A.~Bouguettaya \emph{et~al.}, ``A service computing manifesto: The next 10
  years,'' \emph{Communications of the ACM}, vol.~60, no.~4, pp. 64--72, Mar.
  2017.

\bibitem{8818436}
B.~Shahzaad \emph{et~al.}, ``Composing drone-as-a-service (daas) for
  delivery,'' in \emph{IEEE International Conference on Web Services (ICWS)},
  Milan, Italy, Jul 2019, pp. 28--32.

\bibitem{west2015drone}
G.~West, ``“drone on: The sky’s the limit, if the faa gets out of the
  way,'' \emph{Foreign Affairs}, vol.~94, no.~3, pp. 90--97, May 2015.

\bibitem{10.1007/978-3-030-33702-5_28}
B.~Shahzaad \emph{et~al.}, ``Constraint-aware drone-as-a-service composition,''
  in \emph{International Conference on Service Oriented Computing
  (ICSOC)}.\hskip 1em plus 0.5em minus 0.4em\relax Springer, 2019, pp.
  369--382.

\bibitem{9284135}
B.~Alkouz \emph{et~al.}, ``Swarm-based drone-as-a-service (sdaas) for
  delivery,'' in \emph{IEEE International Conference on Web Services (ICWS)},
  2020, pp. 441--448.

\bibitem{alkouz2020formationbased}
B.~Alkouz and A.~Bouguettaya, ``Formation-based selection of drone swarm
  services,'' in \emph{EAI Mobiquitous Conference}, 2020.

\bibitem{10}
S.J. Kim \emph{et~al.}, ``Drone flight scheduling under uncertainty on battery
  duration and air temperature,'' \emph{Computers \& Industrial Engineering},
  vol. 117, pp. 291 -- 302, Mar. 2018.

\bibitem{en11092204}
D.~Micallef and G.~Van~Bussel, ``A review of urban wind energy research:
  Aerodynamics and other challenges,'' \emph{Energies}, vol.~11, no.~9, pp.
  1--27, Aug 2018.

\bibitem{Citroni_2019}
R.~Citroni \emph{et~al.}, ``A novel energy harvester for powering small uavs:
  Performance analysis, model validation and flight results,'' \emph{Sensors},
  vol.~19, no.~8, p. 1771, Apr. 2019.

\bibitem{7990193}
M.~Khodayar \emph{et~al.}, ``Rough deep neural architecture for short-term wind
  speed forecasting,'' \emph{IEEE Transactions on Industrial Informatics},
  vol.~13, no.~6, pp. 2770--2779, Dec 2017.

\bibitem{10.1145/3287921.3287932}
P.A. Tu \emph{et~al.}, ``Traveling salesman problem with multiple drones,'' in
  \emph{International Symposium on Information and Communication Technology},
  New York, NY, USA, 2018, p. 46–53.

\bibitem{choi2017optimization}
Y.~Choi and P.M. Schonfeld, ``Optimization of multi-package drone deliveries
  considering battery capacity,'' in \emph{96th Annual Meeting of the
  Transportation Research Board}, Washington, DC, USA, Jan. 2017, pp. 8--12.

\bibitem{selecky2013wind}
M.~Seleck{\`y} \emph{et~al.}, ``Wind corrections in flight path planning,''
  \emph{International Journal of Advanced Robotic Systems}, vol.~10, no.~5, p.
  248, Jan 2013.

\bibitem{SHAHZAAD2021335}
B.~Shahzaad \emph{et~al.}, ``Resilient composition of drone services for
  delivery,'' \emph{Future Generation Computer Systems}, vol. 115, pp.
  335--350, 2021.

\bibitem{10.2307/43943662}
K.P. O'Rourke \emph{et~al.}, ``Dynamic routing of unmanned aerial vehicles
  using reactive tabu search,'' \emph{Military Operations Research}, vol.~6,
  no.~1, pp. 5--30, 2001.

\bibitem{DBLP:journals/corr/abs-1902-03506}
A.~{Sanjab} \emph{et~al.}, ``A game of drones: Cyber-physical security of
  time-critical uav applications with cumulative prospect theory perceptions
  and valuations,'' \emph{IEEE Transactions on Communications}, vol.~68,
  no.~11, pp. 6990--7006, 2020.

\bibitem{6916872}
W.T. Hsu \emph{et~al.}, ``Skyline travel routes: Exploring skyline for trip
  planning,'' in \emph{IEEE International Conference on Mobile Data
  Management}, vol.~2, QLD, Australia, Oct 2014, pp. 31--36.

\bibitem{14}
\BIBentryALTinterwordspacing
G.~V{\'a}s{\'a}rhelyi \emph{et~al.}, ``Flight logs of drone swarms,'' Jul 2018.
  [Online]. Available: \url{https://dryad.figshare.com/articles/
  Flight\_logs\_of\_drone\_swarms/6843977/1}
\BIBentrySTDinterwordspacing

\end{thebibliography}

% that's all folks
\end{document}